\documentclass[12pt,preprint]{elsarticle}
\usepackage[T1]{fontenc}
\usepackage[latin9]{inputenc}
\usepackage{geometry}
\usepackage{color}
\usepackage{textcomp}
\usepackage{amsmath}
\usepackage{amssymb}
\usepackage{graphicx}
\usepackage{esint}

\makeatletter

\newcommand{\lyxdot}{.}

\usepackage{verbatim}
\usepackage{textcomp}
\usepackage{dcolumn}
\usepackage{bm}
\usepackage{hhline}
\usepackage{psfrag}
\usepackage{epsfig}
\usepackage{subfigure}

\biboptions{numbers,sort&compress} 

\journal{Physics Letters A}

\makeatother

\begin{document}

\title{Thermal diode made by nematic liquid crystal}

\author{Djair Melo}

\ead{djfmelo@gmail.com}

\author{Ivna Fernandes}

\address{Instituto de Física, Universidade Federal de Alagoas, Av. Lourival
Melo Mota, s/n, 57072-900 Maceió, AL, Brazil}

\author{Fernando Moraes}

\address{Departamento de Física, CCEN, Universidade Federal da Paraíba, Caixa
Postal 5008, 58051-900 , João Pessoa, PB, Brazil}

\address{Departamento de Física, Universidade Federal Rural de Pernambuco,
52171-900 Recife, PE, Brazil}

\author{Sébastien Fumeron}

\address{Institut Jean Lamour, Université de Lorraine, BP 239, Boulevard des
Aiguillettes, 54506 Vandœ{uvre} les Nancy, France}

\author{Erms Pereira\fnref{label2}}

\fntext[label2]{On leave from: Instituto de Física, Universidade Federal de Alagoas,
Av. Lourival Melo Mota, s/n, 57072-900 Maceió, AL, Brazil.}

\address{Escola Politécnica de Pernambuco, Universidade de Pernambuco, Rua
Benfíca, 455, Madalena, 50720-001 Recife, PE, Brazil}

\begin{frontmatter}

\begin{abstract}
This work investigates how a thermal diode can be designed from a
nematic liquid crystal confined inside a cylindrical capillary. In
the case of homeotropic anchoring, a defect structure called escaped
radial disclination arises. The asymmetry of such structure causes
thermal rectification rates up to 3.5\% at room temperature, comparable
to thermal diodes made from carbon nanotubes. Sensitivity of the system
with respect the heat power supply, the geometry of the capillary
tube and the molecular anchoring angle is also discussed. \end{abstract}
\end{frontmatter}


\section{Introduction}

\label{Introduction}

Although the grounds of thermodynamics were formulated by Sadi Carnot
as far back as 1824, controlling heat flows is still a challenging
topic. Besides its academic interest, the worldwide raise of energy
consumption made wasted heat collection and revalorization a major
issue. As a consequence, studies dedicated to thermal management are
now receiving considerable attention \citep{biercuk2002carbon,huang2005aligned}.
Among these, there are studies on designing the thermal analogs of
usual electronic components, such as thermal logic gates \citep{li4}
or thermal transistors \citep{li3,saira,phononics}. Controlled phonon
transport has been obtained with from expensive nanostructured artificial
devices that can perform concentrating, shielding or inverting of
conductive heat flux \citep{Maldovan,phononics2,Narayana,Guenneau2012}.
Thermal rectification generally arises because of an asymmetry of
the system along the direction of heat propagation and non-linear
thermal properties \citep{terrano,li1,zettl,dames,tian2012novel,schmotz2011thermal,:/content/aip/journal/apl/98/8/10.1063/1.3559615}, allowing some
of these devices being understood by classical Fourier-law analysis
\citep{hoff1985asymmetrical,hu2006thermal,dames2009solid,miller2009thermal}.
This is usually achieved either by relying on asymmetric geometries
such as constricted nanotubes \citep{phononics}, or by gradients
of physical properties (e.g. pore density in a nanowire \citep{zettl}).

\begin{figure}[tb]
\begin{centering}
\includegraphics[scale=0.5]{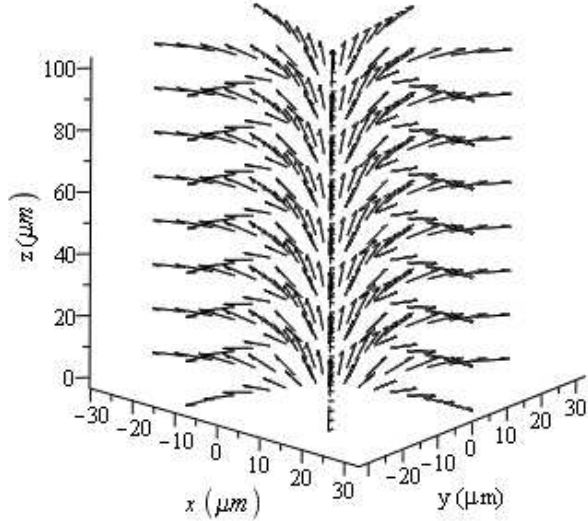} 
\par\end{centering}
\protect\protect\protect\caption{Rod-like molecules forming an escaped radial disclination with escape
in $-\hat{z}$ direction in a capillary tube (not shown) with radius
$R=30\ \mu m$ and height $h=100\ \mu m$. }

\label{fig:ERP} 
\end{figure}

A recent work emphasizes on the assets of nematic liquid crystals
(NLC) for controlling heat flows \citep{cloaking}: contrary to nanostructured
devices \citep{Guenneau2012,qiu,han2014full}, the inner structure
of NLC comes from weak stresses (electric, mechanical...) and/or anchoring
conditions and therefore, the NLC-based device can be tuned at will
for heat conduction. This is possible because NLC are made by anisotropic
molecules that exhibits a stable phase between the crystalline solid
and isotropic liquid phases with, therefore, intermediate properties:
the nematic phase. In such phase, the molecules present orientational
order, but not positional, and they are aligned on average along a
specific direction represented by a unit vector field $\hat{n}\left(\vec{r}\right)$,
known as the \emph{director} field. In particular, the required asymmetry
for the thermal rectification effect can easily be obtained from a
NLC confined inside a capillary tube with homeotropic boundary conditions.
Indeed, as studied by several authors \citep{cladis,Crawford,pieranski},
a funnel-shaped configuration of the director field (known as an escaped
radial disclination) may occur (Fig. \ref{fig:ERP}). The possibility
of manipulation of escaped disclinations by use of laser tweezers
\citep{doi:10.1080/15421400600587787} creates new possibilities for
its use in optical or thermal devices like the one proposed here.

In this paper, we investigate the possibility to design thermal diodes
from confined NLC in cylindrical cavities \citep{Crawford,pieranski,cladis3,burylov,kuzma,PhysRevA.43.6875,PhysRevA.43.835,zografopoulos2006photonic,wang2013polarization,jeong2015chiral,crawford3}
endowed with an escaped radial disclination. Firstly, the thermal
conductivity tensor is determined from algebraic properties of the
director field $\hat{n}\left(\vec{r}\right)$, including the temperature
dependence of the principal molecular thermal conductivities. Secondly,
calculation of the rectification factor using the experimental data
of 5CB \citep{berge} is performed with the following boundary conditions:
a constant temperature is set at one base of the capillary tube, whereas
inward heat flows through the other base. The remainder is thermally
insulated. Reversing that configuration shows a rectification effect.
Finally, we study the influence of the heat power, the geometry and
the anchoring angle on the efficiency of the presented thermal diode,
obtaining rectifications up to 3.5\%, that is higher than the rectification
of thermal diodes made by carbon nanotubes and similar to the ones
made by boron nitride nanotubes \citep{zettl}. The same strategy
here developed can be applied to different anisotropic materials that
present similar orientation field in in a solid state.

\section{Model and formalism}

\label{Model and formalism}

The proposed thermal diode consists of an escaped radial disclination
with escape in the $-\hat{z}$ direction \citep{Crawford,pieranski,crawford3},
Fig. \ref{fig:ERP}. The spatial dependence of the director $\hat{n}\left(\vec{r}\right)$
in the capillary tube is obtained by solving the Euler-Lagrange equation,
with the suitable boundary conditions of this system, of the free
elastic energy density \citep{pieranski} 
\begin{align*}
f= & \frac{1}{2}(\vphantom{}K_{11}\left(div\ \hat{n}\right)^{2}+K_{22}\left(\hat{n}\cdot curl\ \hat{n}\right)^{2}\\
 & +K_{33}\left(\hat{n}\times curl\ \hat{n}\right)^{2}),
\end{align*}
\\
 where $K_{11}$, $K_{22}$ and $K_{33}$ are respectively the curvature
elastic constants of the distortions named \emph{splay}, \emph{twist}
and \emph{bend}. On the one-constant approximation $\left(K_{11}\approx K_{22}\approx K_{33}\equiv K\right)$
and imposing homeotropic condition for the director $\hat{n}$ at
the wall of the capillary tube, we study the rectification related
to the strong and weak anchoring regimes using the algebraic expression
of the spatial director field $\hat{n}\left(\vec{r}\right)$. From
the possible representations of $\hat{n}\left(\vec{r}\right)$ found
in \citep{Crawford,cladis}, we choose the one in \citep{Crawford}
due to its simplicity, such that, for the strong and the weak anchoring
regimes, they are respectively

\begin{equation}
\chi\left(r\right)=2\arctan\left(\frac{r}{R}\right),\label{eq:chi}
\end{equation}
and 
\begin{equation}
\chi\left(r\right)=2\arctan\left(\frac{r}{R}\tan\left(\frac{\chi_{0}}{2}\right)\right),\label{eq:weak}
\end{equation}
where $R$ is the radius of the capillary tube, $\chi\left(r\right)$
is the angle between the director and the axis of the capillary\texttt{{}
}and $\chi_{0}\equiv\chi\left(r=R\right)$ is the anchoring angle.
The physical ideas on energy and their algebraic deduction used on
the obtaining of the eqs. $\left(\ref{eq:chi}\right)$ and $\left(\ref{eq:weak}\right)$
for the escaped radial disclination are found in \citep{Crawford}.
Observe that the algebraic manifestation of the weak anchoring regime
is a pretilt with $\chi_{0}<\frac{\pi}{2}$ \citep{Crawford}.

With $\left(\ref{eq:chi}\right)$ and $\left(\ref{eq:weak}\right)$,
one can write the director of the escaped radial disclination 
\begin{equation}
\hat{n}=\left(\sin\chi\left(r\right),\ 0,\ \cos\chi\left(r\right)\right),\label{eq:disclination-with-escape}
\end{equation}
in cylindrical coordinates and apply it in the equation of the components
of the thermal conductivity tensor \citep{pieranski}, 
\begin{equation}
\lambda^{ij}=\lambda_{iso}\delta^{ij}+\lambda_{a}\left(n^{i}n^{j}-\frac{\delta^{ij}}{3}\right),\label{eq:thermal-conductivity-tensor}
\end{equation}
\\
 where $\delta^{ij}$ are the elements of the homogeneous isotropic
thermal conductivity tensor written in the cylindrical coordinates
and 
\begin{align*}
\lambda_{iso} & =\frac{\lambda_{\|}+2\lambda_{\bot}}{3},\\
\lambda_{a} & =\lambda_{\|}-\lambda_{\bot},
\end{align*}
\\
 being $\lambda_{\|}$ and $\lambda_{\bot}$ respectively the molecular
thermal conductivities parallel and perpendicular to the major molecular
axis, where $\lambda_{\|}>\lambda_{\bot}$. This result with the spatial
asymmetry produced by the escaped radial disclination allow us to
expect that the direction of the ``direct'' thermal setup, the direction
that produces the higher bulk thermal conductivity, is the one of
the escape: $-\hat{z}$ direction. Therefore, our proposed thermal
diode behaves in a similar way to ballistic thermal diodes \citep{miller2009thermal,dames2009solid}. 

Thus the present model could be well developed considering the temperature
independence of the principal thermal conductivities and this simplified
scenario would generate a robustness on the thermal anisotropy of
the escaped radial disclination against the used heat power. However,
as the employed liquid crystal, 5CB, is a thermotropic NLC, the temperature
changes the scalar order parameter and, consequently, their principal
molecular thermal conductivities $\lambda_{\|}$ and $\lambda_{\bot}$.
Considering the influence of the temperature on the scalar order parameter,
such temperature dependences of the thermal conductivities are implemented
using the following nonlinear equations and parameter data extracted
from \citep{berge}: 
\begin{equation}
\begin{split}\lambda_{\|} & =\lambda_{0}+\lambda_{1}\times\left(T-T_{NI}\right)+\lambda_{1,\|}\left(T_{c}-T\right)^{\alpha_{\|}},\\
\lambda_{\bot} & =\lambda_{0}+\lambda_{1}\times\left(T-T_{NI}\right)+\lambda_{1,\bot}\left(T_{c}-T\right)^{\alpha_{\bot}},
\end{split}
\label{temp-conduc}
\end{equation}

where 
\begin{align*}
\lambda_{0} & =1.512,\\
\lambda_{1} & =0.00370,\\
T_{NI} & =308.32\ K,\\
\lambda_{1,\|} & =0.4026,\\
\lambda_{1,\bot} & =-0.1448,\\
T_{c} & =309.21\ K,\\
\alpha_{\|} & =0.237,\\
\alpha_{\bot} & =0.172.
\end{align*}
\\
With the nonlinear behavior of $\left(\ref{temp-conduc}\right)$,
we expect this additional factor facilitates the thermal rectification,
as reported by the literature \citep{hoff1985asymmetrical,dames2009solid,miller2009thermal,dames,li1,phononics,phononics2,li6,hu2006thermal}.

\section{Results and discussion}

\label{Results and discussion}

\subsection{Asymmetric thermal boundary conditions}

\begin{figure}[tb]
\begin{centering}
\includegraphics[scale=0.52]{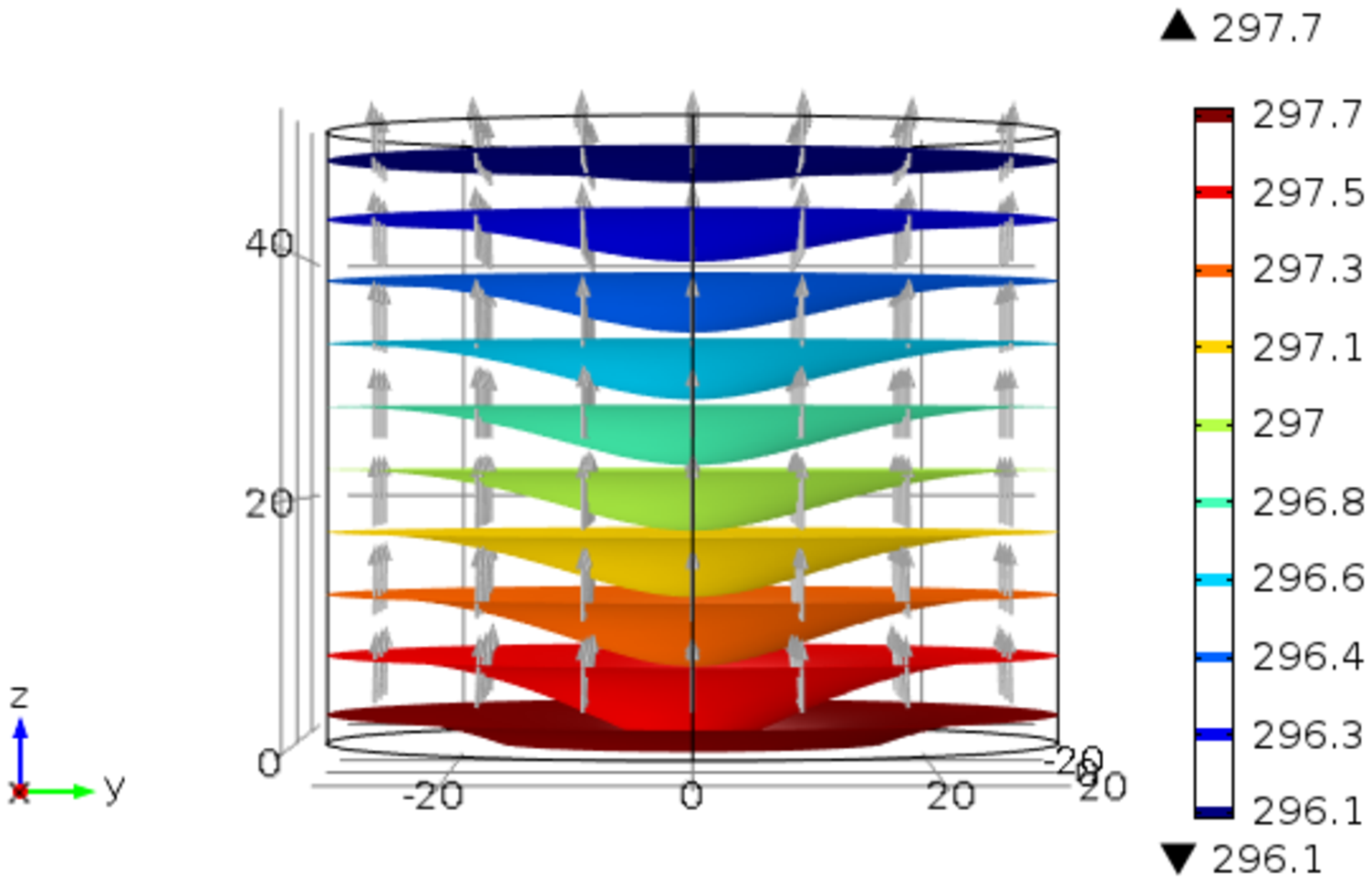} 
\par\end{centering}

\begin{centering}
\includegraphics[scale=0.52]{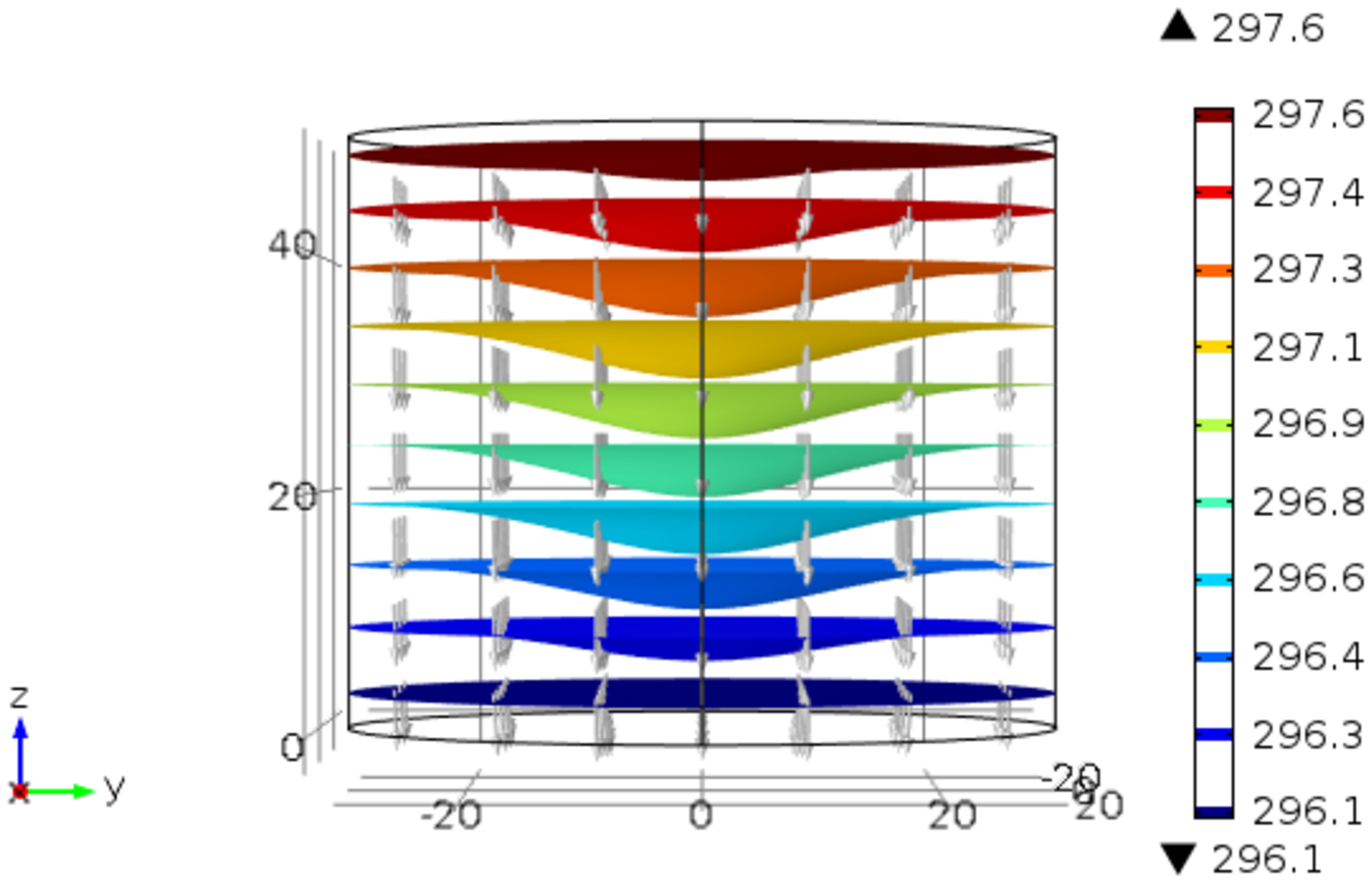} 
\par\end{centering}

\protect\protect\protect\caption{Isothermal surfaces of a liquid crystalline thermal diode in (up)
inverse thermal setup and (bottom) direct thermal setup. The cylinder
has radius $30\ \mu m$, height $50\ \mu m$, inward heat flux of
$5\ \frac{kW}{m^{2}}$ at the base with high temperature and fixed
temperature of $296\ K$ at the other base. The rectification effect
is 1.5\%.}

\label{fig:isothermal-surfaces} 
\end{figure}

We now implement, on the strong anchoring regime with the angle $\chi\left(r\right)$ between
the director and the plane $x-y$ given by $\left(\ref{eq:chi}\right)$,
an escaped radial disclination in the $-\hat{z}$ direction inside
a capillary tube with radius $R=30\ \mu m$ and height $h=100\ \mu m$
in a finite element simulation using COMSOL Multiphysics, Fig. \ref{fig:ERP}.
The whole tube is set at initial temperature of 296 K. One of the
bases of the cylinder is set at 296 K, heat is pumped in at a rate
of $5\ \frac{kW}{m^{2}}$ across the other one and thermal insulation
is set at the lateral surface.
These information are implemented in our simulation to solve, in
steady-state regime and in the absence of internal sources, the Fourier
and Laplace equations \citep{ozisik} 
\begin{align*}
q^{i} & =-\lambda^{ij}\partial_{j}T,\\
\nabla\cdot\vec{q} & =\partial_{i}\left[\lambda^{ij}\partial_{j}T\right]=0,
\end{align*}
where $q^{i}$ are the components of the heat flux $\vec{q}$ and
$T$ is the temperature field. The results are shown on Fig. \ref{fig:isothermal-surfaces}.

To determine the rectification, we use the following definition: 
\begin{equation}
Rectification=\frac{\Delta T_{i}-\Delta T_{d}}{\Delta T_{d}}\ \times100\ \ \ \ \%,\label{eq:rectification}
\end{equation}
where $\Delta T_{i}=T_{i,h}-296$ is the difference between the high
temperature produced by the heat pumped in the cylinder, $T_{i,h}$,
when working in the inverse setup and the low temperature of 296 K
(similarly, we have $\Delta T_{d}=T_{d,h}-296$ when working in the
direct setup).

From the simulations of the capillary tube, represented in the Fig.
\ref{fig:isothermal-surfaces}, we observed that the downward flux
produces the lower difference on the temperatures, meaning that this
direction has the higher thermal conductance, representing the direct
setup of the thermal diode. With the values of $\Delta T_{i}$ and
$\Delta T_{d}$ of this capillary tube, the eq. $\left(\ref{eq:rectification}\right)$
produces a rectification effect of $1.5\%$, similar to the one produced
by carbon nanotubes \citep{zettl} at room temperature. This rectification
can be explained considering the spatial asymmetry produced by the
escape in $-\hat{z}$ direction, in a similar way to ballistic rectification
\citep{miller2009thermal,dames2009solid} due to spatial asymmetries,
or considering the addition of an effective curvature felt by the
phonons on such direction \citep{cloaking,fumeron1}, once $\lambda_{\|}>\lambda_{\bot}$,
promoting a higher thermal conductivity in the direction of the escape.

\subsubsection{Efficiency of the thermal diode}

\label{Efficiency of the thermal diode}

We analyzed the efficiency of the rectification due to the heat power
supply, to the geometry of the capillary tube and to the molecular
anchoring angle at the surface of the tube. The heat powers were chosen
to produce temperatures bellow the nematic-isotropic transition temperature
$T_{NI}$ in eq. $\left(\ref{temp-conduc}\right)$, while the radii
of the capillary tube were chosen to be far bigger than a typical
value of the core radius $r_{c}$ of a radial disclination ($r_{c}\approx500\ \text{\AA}$
\citep{pieranski}), avoiding energy competition between the radial
disclination and the escaped radial one \citep{cladis}. The upper
and lower limits for the height were assumed respecting experimental
issues on the creation of the escaped radial disclination in the capillary
tube (for example, the flow profile of the liquid crystal during the
filling of the capillary tube \citep{pieranski} or the temperature
rate during the cooling of the liquid crystal from the isotropic phase
to the nematic one \citep{Crawford}).

Relative to the heat power and to the geometry of the capillary tube,
the results are summarized in Fig. 3. It was observed that the rectification
effect decreases with the heat power. Its reason is that more heat
power implies in a higher temperature, decreasing the anisotropy $\lambda_{\|}-\lambda_{\bot}$
, as can be seen by the equations $\left(\ref{temp-conduc}\right)$,
reducing the rectification effect.

\begin{figure}[t]
\begin{centering}
\includegraphics[scale=0.54]{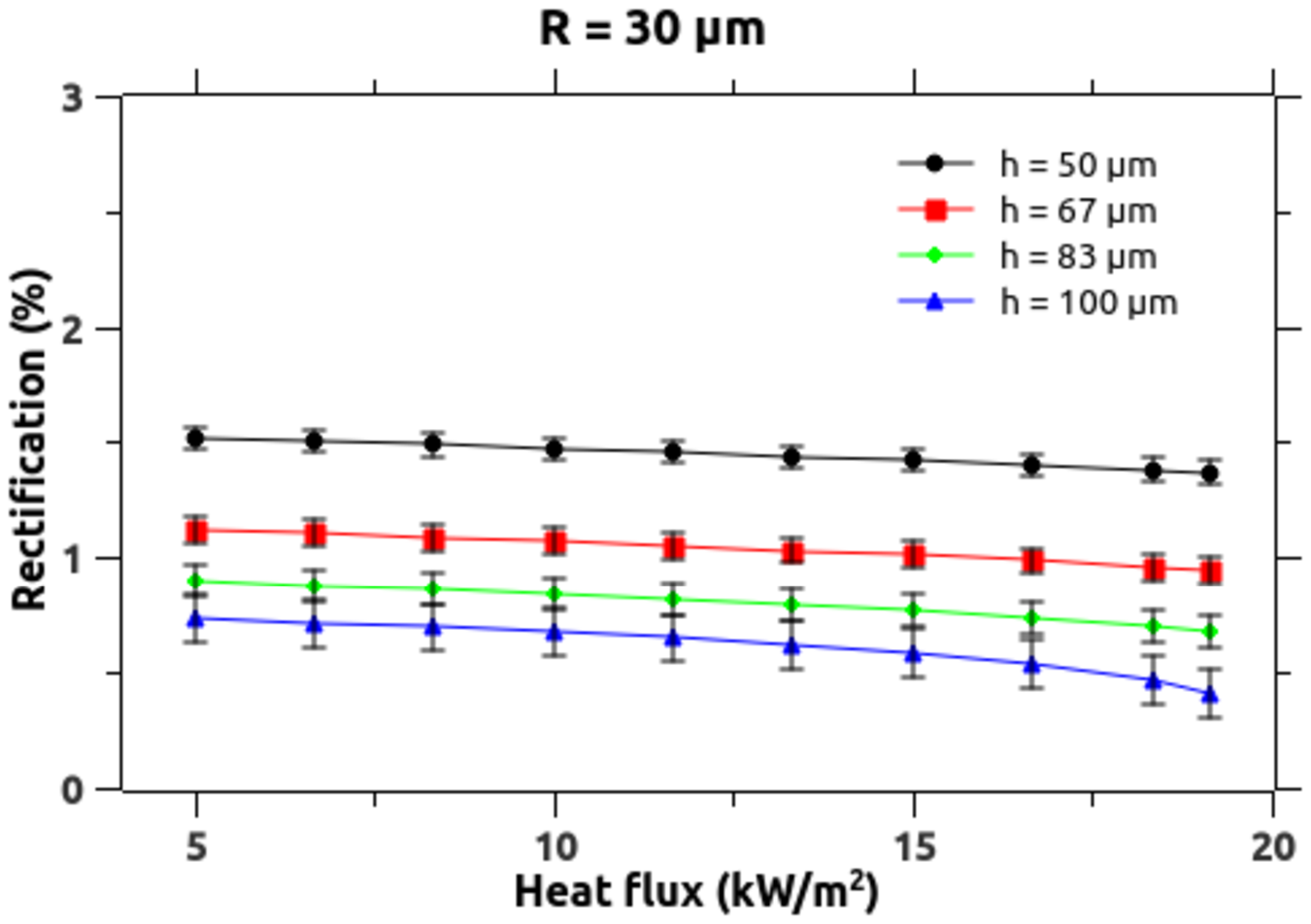}\includegraphics[scale=0.54]{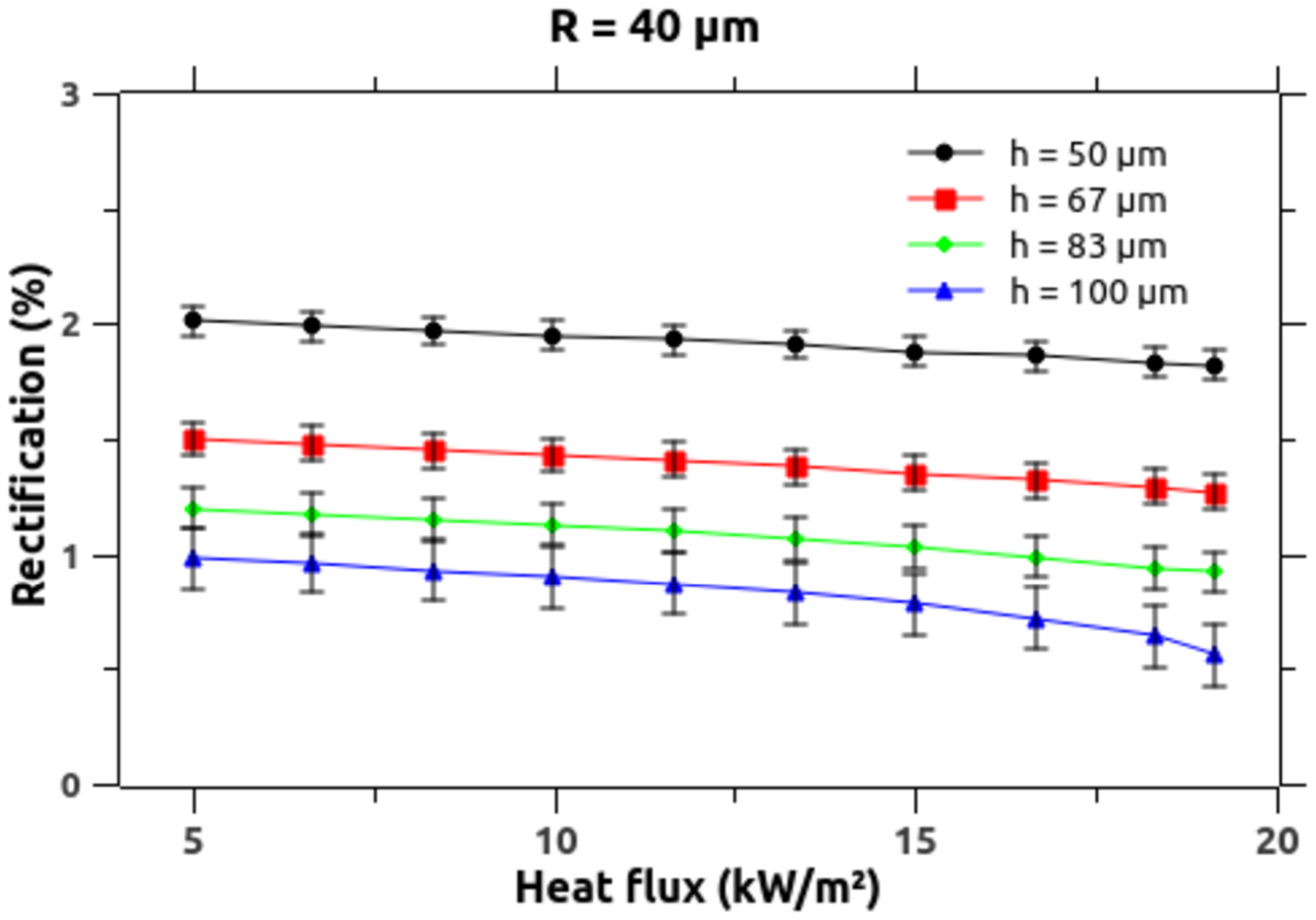} 
\par\end{centering}

\begin{centering}
\includegraphics[scale=0.54]{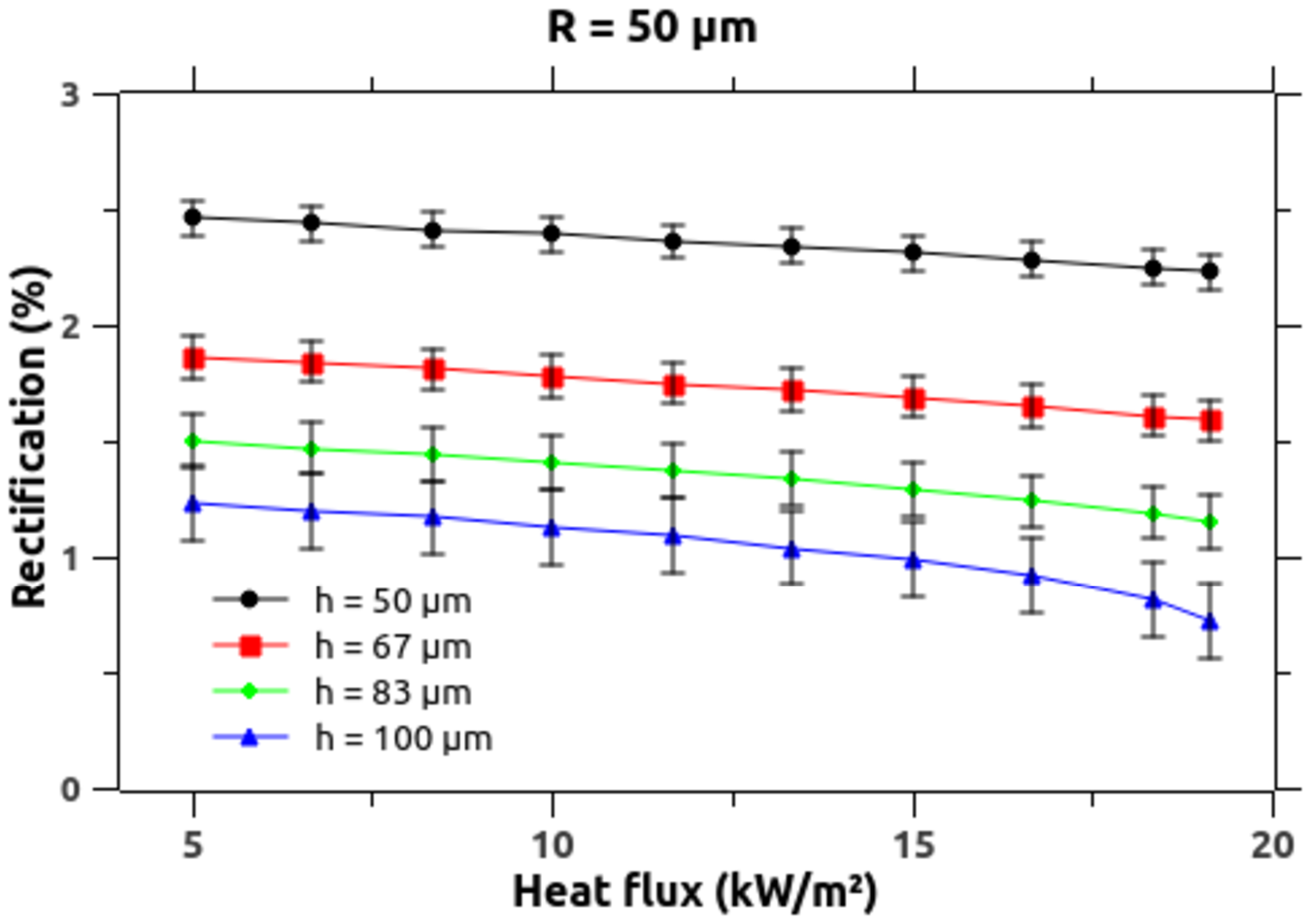}\includegraphics[scale=0.54]{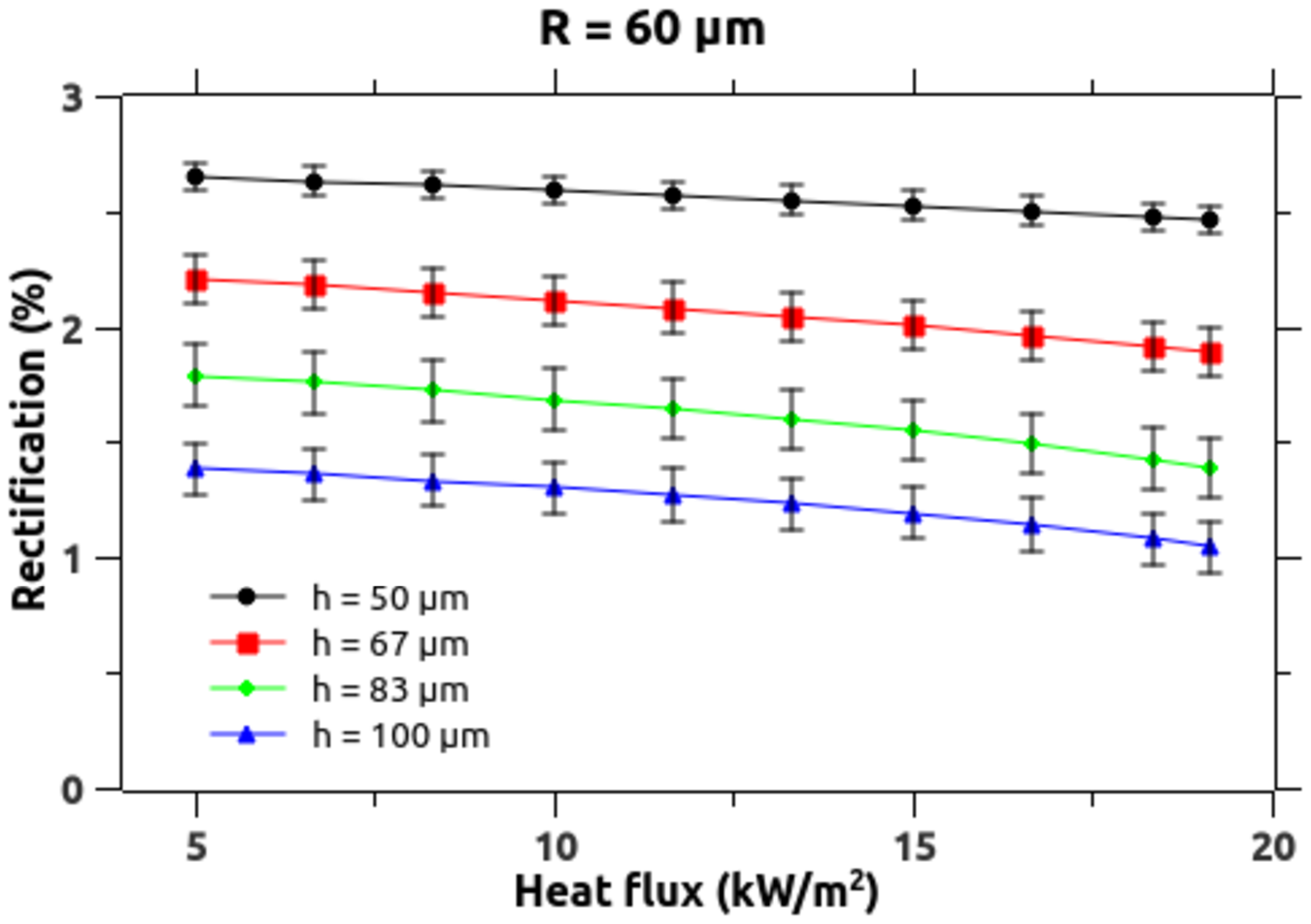} 
\par\end{centering}

\begin{centering}
\label{fig:geo-analysis} 
\par\end{centering}

\caption{Rectification versus heat power produced by an escaped radial disclination
in the strong anchoring regime, eq. $\left(\ref{eq:chi}\right)$,
with different values of the radius $R$ and the height $h$ of its
confining capillary tube.}
\end{figure}

Relative to the geometry, it was observed in Fig. 3 that the rectification
effect increases with increasing area of the bases and increases with
decreasing height of the tube. About the influence of the area on
the rectification, we justify the results noting that a large area
denotes more heat flux to be deflected by the isothermal surfaces,
enhancing the rectification. About the influence of the height on
the rectification, the results suggest that when $h/R$ is high, the
device looks like a 1D wire, minimizing the effect of the escaped
radial disclination on the heat flux, decreasing the rectification.
Such dependence on the height favors the miniaturization of the diode,
while the dependence on the area concurs to the application of this
device on a large region, in opposite of its use on a small local.

\begin{figure}[!tb]
\begin{centering}
\includegraphics[scale=0.8]{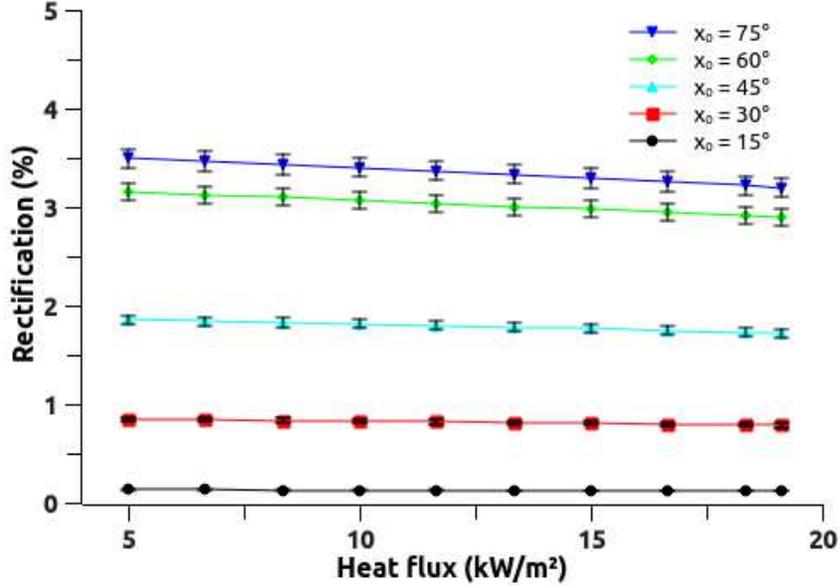} 
\par\end{centering}

\protect\caption{Rectification versus heat power produced by an escaped radial disclination
in the weak anchoring regime, eq. $\left(\ref{eq:weak}\right)$, inside
a capillary tube with radius $R=60\ \mu m$, height $h=50\ \mu m$,
with different anchoring angles $\chi_{0}<90\text{\textdegree}$.}

\label{fig:anchoring} 
\end{figure}

Relative to the anchoring angle, we use eq. $\left(\ref{eq:weak}\right)$
in the simulations, where the rectification for different values of
the anchoring angle $\chi_{0}<90\text{\textdegree}$ are found in
Fig. \ref{fig:anchoring} with radius $R=60\ \mu m$ and height $h=50\ \mu m$,
obtaining a rectification up to 3.5\%, that is higher than the rectification
obtained by carbon nanotubes and comparable to the ones made by boron
nitride nanotubes \citep{zettl}. Comparing this value with the results
in Fig. 3, we notice an increased rectification with $\chi_{0}=75\text{\textdegree}$
and $\chi_{0}=60\text{\textdegree}$ and a monotonically decrease
from $\chi_{0}=75\text{\textdegree}$ to $\chi_{0}=15\text{\textdegree}$.
The behavior is explained by a reduction on the anisotropy of the
capillary tube and, consequently, on the effective curvature felt
by the phonons \citep{pereira2013metric} and by the heat flux \citep{fumeron1},
once the spatial configuration of the director $\hat{n}$ approaches
to the uniform $\hat{n}=\hat{z}$. This also implies the existence
of an optimized anchoring angle $\chi_{0}$ that generates the maximum
rectification. In fact, when we reduce the anchoring angle $\chi_{0}$,
reducing the axial anisotropy, it is expected that there is an ideal
anchoring angle $\chi_{ideal}$ that maximizes the rectification.
Below this $\chi_{ideal}$, we obtain molecular configurations that
makes the rectification decreases. This behavior is shown in Fig.
5, where the ideal anchoring angle is $\chi_{ideal}=73\text{\textdegree}$.
For anchoring angles smaller than $\chi_{ideal}=73\text{\textdegree}$,
the anisotropy starts to be destroyed, causing the thermal diode becomes
uniform throughout the capillary tube.

\begin{figure}[!tb]
\begin{centering}
\includegraphics[angle=-90,scale=0.6]{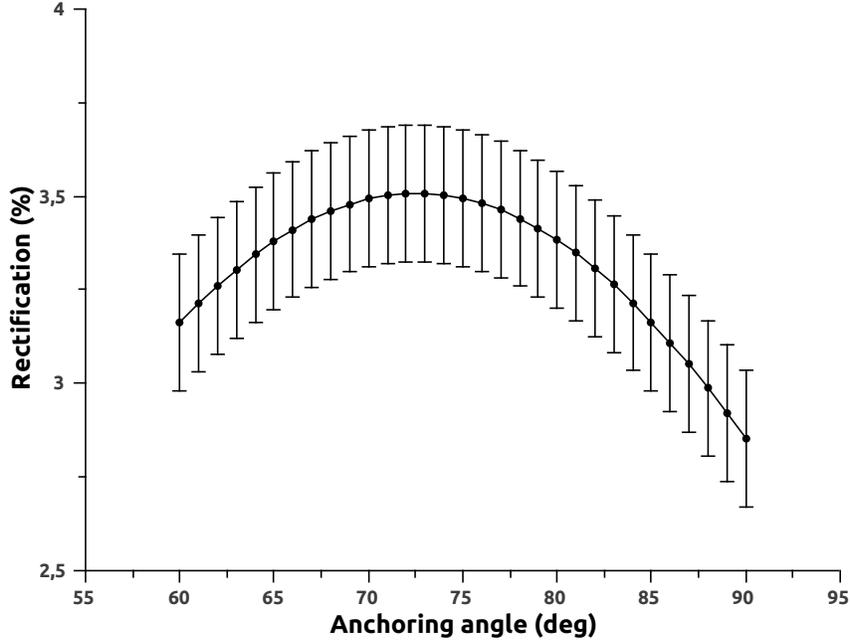} 
\par\end{centering}

\label{fig:ret-angles}

\caption{Rectification versus anchoring angles $\chi_{0}$ produced by an escaped
radial disclination in the weak anchoring regime inside a capillary
tube with radius $R=60\ \mu m$, height $h=50\ \mu m$ and heat flux
$5\ \frac{kW}{m^{2}}$.}
\end{figure}

For different geometries and anchoring regimes, our simulations resulted
in temperatures near to $T_{NI}=308.32\ K$ at heat fluxes from $19\ \frac{kW}{m^{2}}$,
for radius $30\ \mu m$, height $100\ \mu m$ and anchoring angle
$\chi_{0}=0$, up to $37\ \frac{kW}{m^{2}}$, for radius $60\ \mu m$,
height $50\ \mu m$ and anchoring angle $\chi_{0}=15\text{\textdegree}$.
These results on heat flux, bigger than other reported one in \citep{tian2012novel},
are explained due to their rectifications (using $5\ \frac{kW}{m^{2}}$
, 0.73\% for the first case and 3.5\% for the latter one), because
high rectifications presented low temperatures and, consequently,
strong robustness for the heat flux. Thus eutectic mixtures could
enlarge the heat power range and the temperature range, avoiding the
loosing the desired anisotropy.

\begin{figure}
\begin{centering}
\includegraphics[scale=0.54]{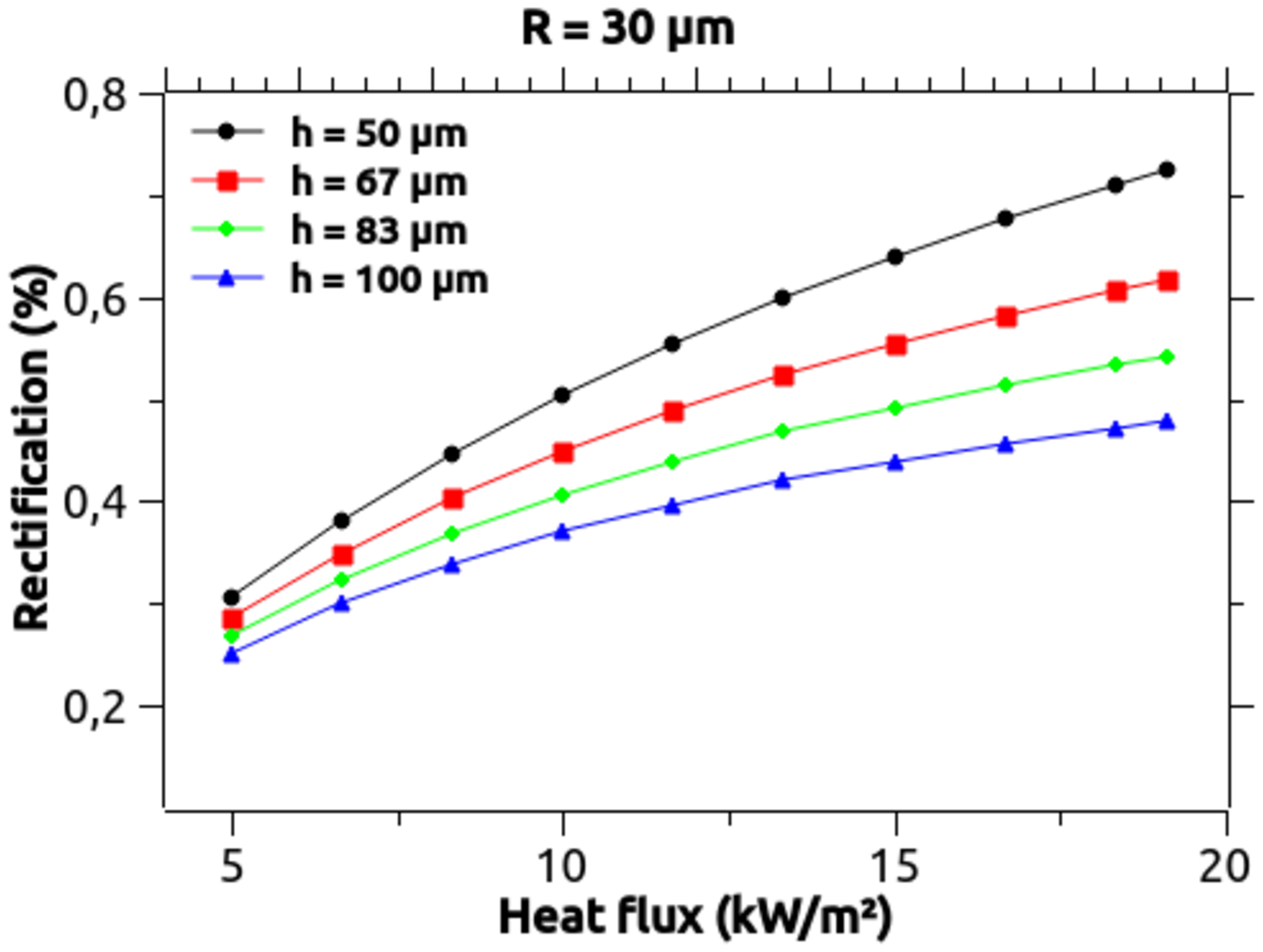}\includegraphics[scale=0.54]{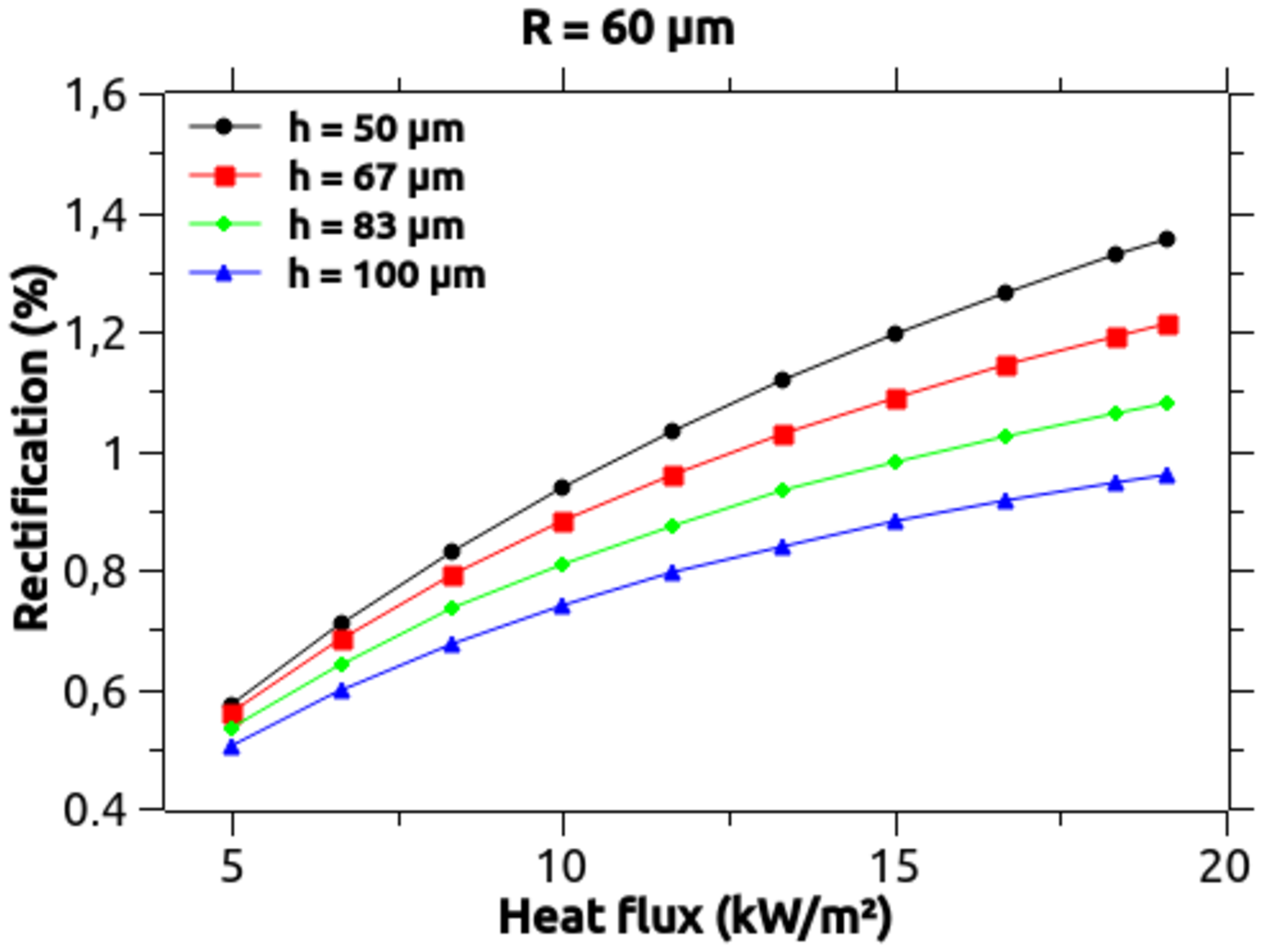}
\par\end{centering}

\caption{Rectification versus heat power produced by an escaped radial disclination
with constant $\lambda_{\|}$ and $\lambda_{\bot}$ in the strong
anchoring regime, eq. $\left(\ref{eq:chi}\right)$, with different
values of the radius $R$ and the height $h$ of its confining capillary
tube. The errors on the rectification range from $10^{-4}$ to $10^{-7}$.}

\label{fig:constant-lambda}
\end{figure}

To isolate the asymmetric aspect from the nonlinear one of the rectification,
we repeat the simulations disregarding the temperature dependence
of $\lambda_{\|}$ and $\lambda_{\bot}$ and set them constant at
299.26 K \citep{berge}. The results are shown in Fig. \ref{fig:constant-lambda}.
As can be seen, the constant molecular thermal conductivities produce
rectifications that are proportional to the heat flux and that are
low compared to Fig. 3. This shows that the nonlinear temperature
dependences of the $\lambda_{\|}$ and $\lambda_{\bot}$ has strong
influence on the rectification and the studied spatial anisotropy
is enough for such phenomenon. The presence of the rectification due
to the asymmetry can be understood by the tendency of the phonon to
follow the director \citep{pereira2013metric,fumeron1}, once $\lambda_{\|}>\lambda_{\bot}$
, while the proportionality to the heat flux can be explained by the
available heat to be deflected by isothermal surfaces, similar to
the influence of the radius $R$. 

\subsection{Symmetric thermal boundary conditions }

\begin{figure}[tb]
\begin{centering}
\includegraphics[scale=0.53]{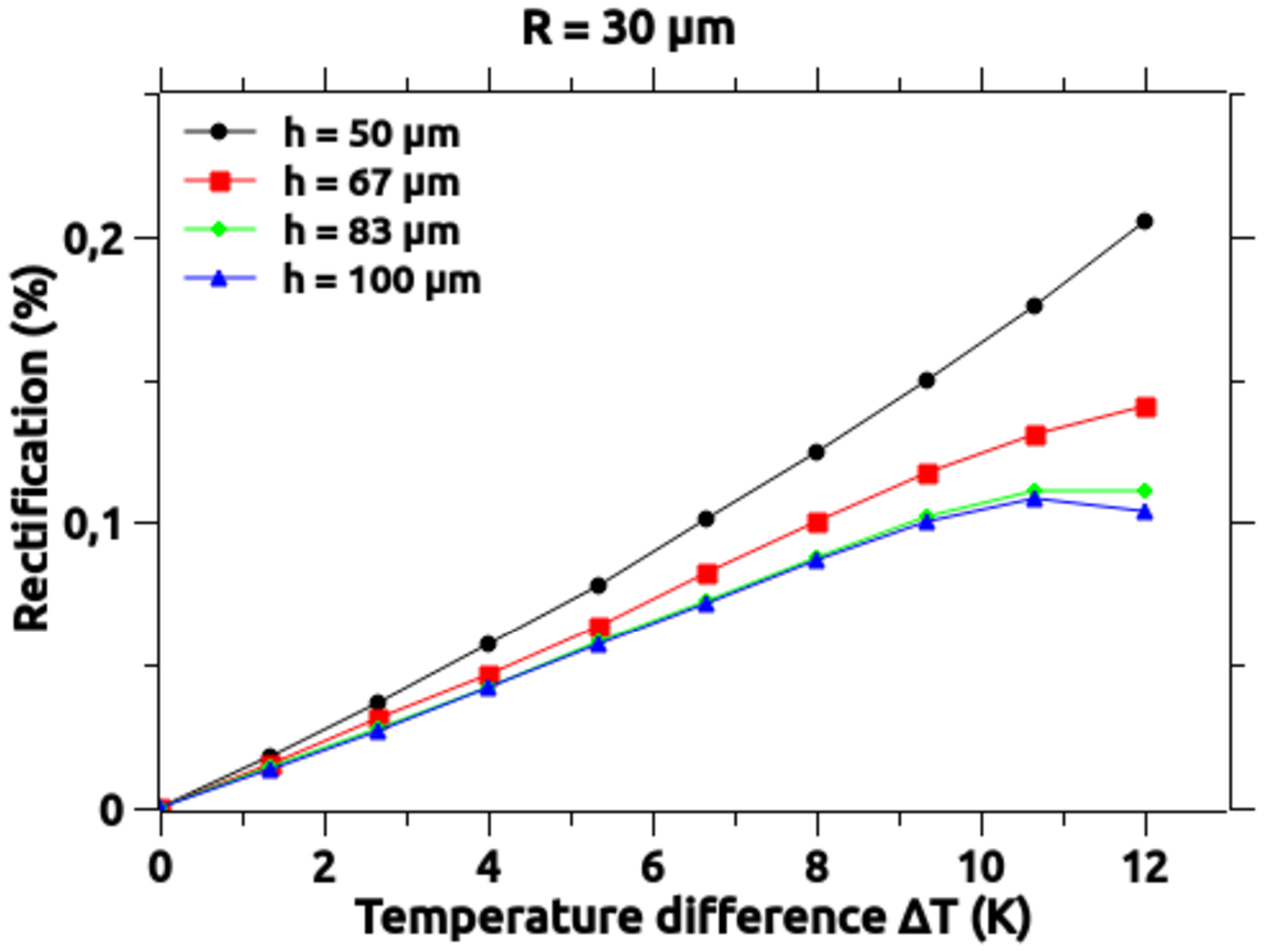}$\ $\includegraphics[scale=0.53]{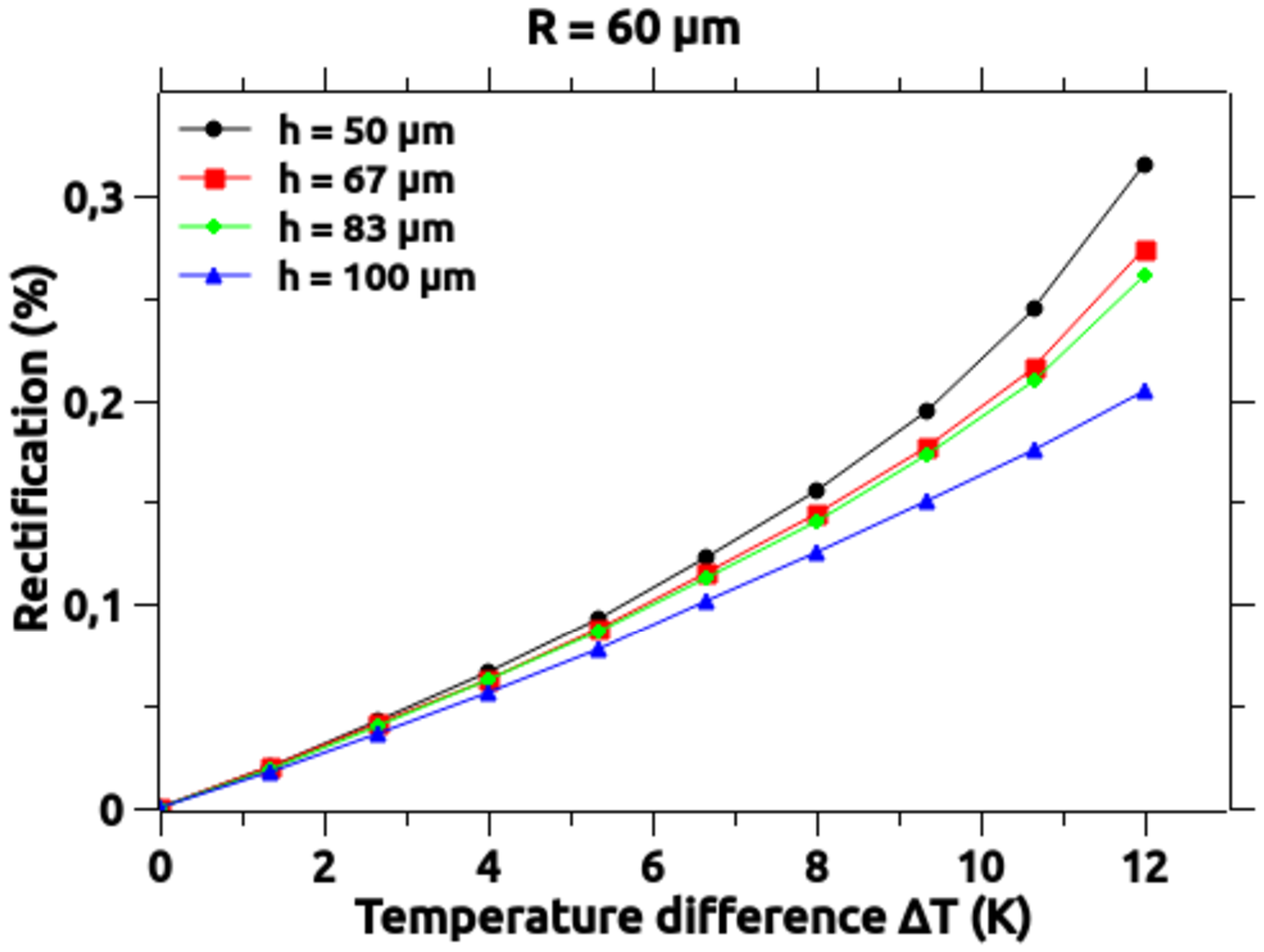}
\par\end{centering}

\caption{Rectification versus temperature difference$\Delta T\equiv T_{high}-296$,
with $T_{high}\in\left[296\ K,\ 308\ K\right]$, produced by an escaped
radial disclination in the strong anchoring regime, eq. $\left(\ref{eq:chi}\right)$,
with different values of the radius $R$ and the height $h$ of its
confining capillary tube. The errors on the rectification range from $10^{-4}$ to $10^{-7}$.}

\label{fig:temp-bias-temp-depend}
\end{figure}

Using again eq. $\left(\ref{temp-conduc}\right)$ and for symmetric
thermal boundary conditions, where on both basis are set constant
temperatures producing a temperature difference $\Delta T\equiv T_{high}-296$
with $T_{high}\in\left[296\ K,\ 308\ K\right]$, we found the rectifications
shown in Fig. \ref{fig:temp-bias-temp-depend}, where it was used
another definition for it \citep{dames2009solid}: 
\begin{equation}
Rectification=\frac{Q_{d}-Q_{i}}{Q_{i}}\ \times100\ \ \ \ \%,\label{eq:rectification-1}
\end{equation}
\\
where $Q_{d}$ and $Q_{i}$ $\left(Q_{d}>Q_{i}\right)$ are the heat
flux in the direct (with $T\left(x=h\right)=T_{high}$ and $T\left(x=0\right)=296\ K$)
and inverse (with $T\left(x=0\right)=T_{high}$ and $T\left(x=h\right)=296\ K$)
setups respectively. From Fig. \ref{fig:temp-bias-temp-depend} we
notice that the rectifications are proportional to the temperature
bias set on the basis (unless of the last points for $R=30\ \mu m$
and $h=83\ \mu m$ and $h=100\ \mu m$), consistent to \citep{hoff1985asymmetrical,hu2006thermal,dames2009solid},
and are very low compared to the results from the asymmetric boundary
conditions. This suggests our thermal diode based on liquid crystal
works better as a heat flux thermal diode, when one sets a heat flux
at one the basis, than a temperature bias thermal diode, when one
sets constant temperatures on the basis. As shown in \citep{hoff1985asymmetrical},
Neumann theorem forbids asymmetrical transport if the inner structure
is symmetrical and if the thermal conductivity is temperature independent.
The fact that we have neither, allows us to have a greater asymmetry
in the heat flux when the boundary conditions change form temperature
bias to heat injection. We can also think of this in terms of phonon
beams: depending on the way they traverse the device, they are either
focused or defocused due to the asymmetic director field configuration.
This effect has been observed experimentally for light beams in escaped
disclinations, as reported in \citep{doi:10.1117/12.2079096}. This
can indicate that the phonons that start to move from the fixed higher
temperature to the fixed lower one fell less spatial anisotropy of
the thermal conductivity than the phonons that already are in motion
due to the external heat flux set on one of the terminals of the diode.\textbf{
}We also notice in Fig. \ref{fig:temp-bias-temp-depend} for $R=30\ \mu m$
a decreasing on the rectification for $\Delta T=T_{high}-296\approx12\ K\Rightarrow T_{high}\approx308\ K.$
This can be understood as a competition between the available heat
flux due the $\Delta T$ and the annihilation of the anisotropy between
the molecular thermal conductivities $\lambda_{\|}$ and $\lambda_{\bot}$near
to the nematic-isotropic phase transition $T_{NI}$.

\section{Summary and conclusions}

\label{Summary-conclusions}

\noindent The present paper continues recent studies \citep{cloaking,fumeron1}
that explore the possibilities offered by NLC for heat transfer management
and their technological applications for phononics. We proposed a
thermal diode made of nematic liquid crystal, that consists in an
escaped radial disclination of the liquid crystal 5CB in a capillary
tube. Once the parallel molecular thermal conductivity of the NLC
is higher than the perpendicular one, the direction that the thermal
diode conducts (the direct thermal setup) is on the escape of the
disclination. For different values of the radius and height of the
tube and the anchoring angles at the capillary walls, rectification
effects are observed in our simulations up to 3.5\% at room temperature.
Increasing heat power and the height of the tube decrease the rectification
effect, while increasing the area of the tube increases this phenomenon.
Such dependence on the geometry would lead to miniaturization of the
device and its application on a vast surface. We found that our thermal
diode works better as a heat flux thermal diode, pumping heat in the
diode, than a temperature bias thermal diode, setting a fixed difference
temperature on the terminals of the diode. We also noticed that the
spatial anisotropy alone of the thermal conductivity produces the
thermal rectification for our proposed thermal diode, with low rectifications
when compared with the ones produces by the spatial asymmetry plus
temperature non-linearities. 

The results presented in this article were based on four factors:
a physically anisotropic molecular arrangement, temperature dependence
of the thermal conductivities, the existence of a nematic phase and
the possibility of elastic deformations (topological defects) in the
nematic phase. Thus, our results apply to lyotropic liquid crystals,
once they have three of these factors \citep{cuppo2002thermal,mccormick2003photopolymerization},
and any other materials that sustains such director configurations
in a nematic phase, as ferronematics \citep{matuo2002lyotropic,satiro2009light,zadorozhnii2006nematic}.
With the prospect of having glassy liquid crystals \citep{van2007glassy,yeap2008novel},
it is possible to release the escaped radial disclination from its
capillary tube, enlarging the technological applications of the presented
thermal diode.

A natural extension of this work is to design thermal transistors
\citep{li3,saira,phononics} from liquid crystals. The idea is to
explore the non-linearity of the principal thermal conductivities
in Fourier's equation to find thermally stable states that \textquotedbl{}cut\textquotedbl{}
and \textquotedbl{}release\textquotedbl{}' the heat flow or that amplify
it.

\section*{Acknowledgements}

\label{Acknowledgements}

F.M. and E.P. acknowledges CAPES, CNPq, and E.P. to FAPEAL (Brazilian
agencies) for the financial support. The authors thank the anonymous
referees for the comments.


\label{References}

\bibliographystyle{elsarticle-num}
\bibliography{/home/djair/Dropbox/Thermal_Diode/ARTICLE/Thermal_Diode_PLA/ERMS-Thermal_Diode_PLA-reviewers/pereira.bib}

\end{document}